\documentclass{article}
\usepackage[english]{babel}

\usepackage[a4paper,top=2cm,bottom=2cm,left=2.5cm,right=2.5cm,marginparwidth=1.75cm]{geometry}
\usepackage{mathrsfs}
\usepackage{amsmath}
\usepackage{amssymb}
\usepackage{graphicx}
\usepackage{siunitx}
\usepackage{caption}
\usepackage{subcaption}
\usepackage{authblk}
\usepackage{multirow}
\usepackage{setspace} 
\usepackage{color}
\usepackage{soul}
\usepackage[colorlinks=true, allcolors=blue]{hyperref}

\begin{document}

\title{Damage Prediction of Sintered $\alpha$-SiC Using Thermo-mechanical Coupled Fracture Model}
\author{Jason Sun, Yu Chen, Joseph J. Marziale, Eric A. Walker, David Salac, James Chen$^*$}
\affil{Department of Mechanical and Aerospace Engineering, The State University of New York at Buffalo, Buffalo, New York 14260, United States\\$^*$Author to whom correspondence should be addressed. e-mail: chenjm@buffalo.edu}
\date{October 2022}

\maketitle

\begin{abstract}
A three-way coupled thermo-mechanical fracture model is presented to predict the damage of brittle ceramics, in particular $\alpha$-SiC, over a wide range of temperatures (20-1400$^\circ$C). Predicting damage over such a range of temperatures is crucial for thermal protection systems for many systems such as spacecraft. The model, which has been implemented in MOOSE, is divided into three modules: elasticity, damage phase field, and heat conduction. Analytical approaches for determining crack length scales are presented for both simple tension and simple shear. Validation tests are conducted for both flexural strength and fracture toughness over the specified range of temperatures. Flexural strength simulation results fall within the uncertainty region of the experimental data, and mode I fracture toughness simulation results are also in agreement with the experimental data. Mode II and mixed mode fracture toughness simulations results are presented with the modified G-criterion. Finally, the parallel computing capabilities of the model is considered in various scalability tests. \\

\end{abstract}

\section{Introduction}
The idea of integrating the processing and structure of materials with the target goals and properties is not new. The concept of an Integrated Computational Materials Engineering (ICME) approach, whereby computational tools are utilized to achieve this goal of integration, is newer and has been introduced by Olson~\cite{Olson1997, Olson1998}, and has also been known as Through Process Modelling~\cite{Hirsch2006}. The concept behind ICME is to use a systematic approach to facilitate the cycle between commercial design and production, with a linear hierarchical relationship among processing, structure, properties, and performance, as shown in Fig. 1~\cite{Olson1997}. All of the internal connections are vital in material design processes.

While substantial research effort exists on developing ICME for alloys exist~\cite{Horstemeyer2012, Zhang2021, Wang2020, Raturi2019, Taylor2018}, there are much fewer studies on ICME for advanced ceramics. The recent and increasing maturity of advanced ceramics has drawn more attention in research and industrial applications~\cite{Raether2018}, which has increased the need for true ICME efforts for ceramics that can bridge models and experimental data to accelerate material developments for applications such as those required for spacecraft systems. Examples include Shi et al, where it was proposed to use an ICME approach for Ceramic Matrix Composite Development with a focus on aircraft engine applications using SiC/SiC composites~\cite{Shi2012}. Raether et al gives an overlook of ICME for ceramics with examples of Zirconia Toughened Alumina ceramics~\cite{Raether2018}. Raether's work uses defects obtained from scanned images as a mesh input to calculate stress concentration{\color{black}. However, study in} fracture strength {\color{black} of ceramics through ICME is still lacking. In particular, fracture or damage under high temperatures or hypersonic conditions is in the critical need for aerospace applications.}

Material modeling plays an important role in predicting the relationship between material properties and performance, with engineering designs often requiring the ability to predict where and when cracks will appear over a wide range of temperatures. Various attempts have been made to model the fracture mechanics of brittle materials such as ceramics. Miehe et al first proposed a phase field model that separated the elastic strain into tensile and compressive parts, where damage only affects the tensile portion~\cite{Miehe2010}. Clayton et al proposed a phase field model that captures both the micro-structure and fracture of heterogeneous poly-crystal ceramic composites~\cite{Clayton2019}. Wu et al also showed that phase field modeling performs well for brittle materials~\cite{Wu2018}. {\color{black} Zuo et al proposed a phase field model to predict thermal-shock induced fracture of Al$_2$O$_3$ and 3Y–ZrO$_2$ through water quenching, but mechanical loading was not considered~\cite{Zuo2022}. Similarly, Chu el al used phase field method to model the dynamic crack path of brittle material subjected to thermal shock~\cite{Chu2017}. Ruan et al also utillizied a thermo-mechanical phase field model to model the defects induced by solidification shrinkage in additive manufacturing \cite{Ruan2023}. }While there has been rapid growth of material modeling efforts regarding ceramics, none of these studies provides ceramic damage prediction under a wide range of temperatures.

The simulation model used in this work is a coupled scheme involving fracture mechanics, elasticity, and heat conduction. Fracture mechanics is arguably the most important engineering consideration for materials. Linear elastic fracture mechanics (LEFM) assumes that the stressed material is brittle and undergoes little plastic deformation before fracture, which is an appropriate assumption for ceramics. In general there are two ways to predict damage: (1) the sharp interface model, where tracking the complex crack surface is required, and (2) a diffuse crack model, which approximates the crack interface as a diffusive zone. In the latter case, which is the basis of this work, an auxiliary scalar function provides a continuous description of the damaged state at any location and time. Griffith's fracture theory was the first of its kind to use energy criteria, positing that the energy required to create new crack surfaces is equivalent to the stored strain energy being released during crack propagation~\cite{Lajtai1971}. This concept is used to define crack initiation and propagation criteria. The phase field model is used to capture the crack evolution and propagation using the variational principle of the free energy. For a non-conserved phase field variable, the Allen-Cahn equation is utilized to describe this reaction-diffusion system~\cite{Allen1972}. A heat conduction model is also included to account for temperature effects.

The primary objective of this study is to implement the thermal-mechanical coupled fracture model for damage prediction {\color{black} over a wide range of temperatures}. The framework is built in the open source Multiphysics Object Oriented Simulation Environment (MOOSE)~\cite{Permann2020}. The ceramic $\alpha$-SiC is used as an exemplar to show that this proposed framework is capable of modeling fracture phenomena over a wide range of temperatures. The material $\alpha$-SiC is chosen due to it's many applications, from abrasives to semiconductors~\cite{Ma2022, Matsunami2020}. Because of its high wear resistance and strength at high temperatures, $\alpha$-SiC is also a leading candidate for use in thermal protection systems~\cite{Al-Jothery2020}. Therefore the proposed framework is designed with a view towards $\alpha$-SiC applications. The rest of this report is organized as follows: the three different modules and how they are coupled are presented in Sec.~\ref{theorysection}; validation tests that compare simulation results with experimental data are shown in Sec.~\ref{validation section}; and the parallel computing capacity of the multiphysics model is considered in Sec.~\ref{scallability section}.

\section{Theory} \label{theorysection}
The problem at hand is to build a framework to accurately model the fracture phenomena of $\alpha$-SiC ceramics for the high temperatures. The framework presented here contains three modules: elasticity, phase field, and heat conduction, each of which are discussed in detail below.

\subsection{Elasticity}
Elasticity governs the mechanical behavior of the material. The model assumes linear elasticity, since ceramics are brittle in nature, and any body forces are negligible. An isotropic assumption is applied as there is no directionality in material properties. Also, as displacement occurs at a small time scale, a quasi-steady state is assumed. The equation of motion {\color{black} at equilibrium }is given by
\begin{equation}
    \sigma_{ij,j} = 0, \label{equationofmotion}
\end{equation}
where $\sigma_{ij}$ is the Cauchy stress tensor and indices $i,j$ subsume directions $x, y, z$. With the small strain assumption, the elastic strain is 
\begin{equation}
    \varepsilon^E_{ij} = \frac{1}{2}(u_{i,j}^E + u_{j,i}^E),
\end{equation}
where $u_{i,j}^E$ represents the elastic displacement gradient. By definition, the stress is equal to the partial derivative of elastic strain energy with respect to elastic strain; i.e.,
\begin{equation}
    \sigma_{ij}^{E} = \frac{\partial f_{elas}}{\partial \varepsilon_{ij}^E} \label{felas_eij}.
\end{equation}
For isotropic materials, the elastic strain energy 
\begin{equation}
    f_{elas} =  \omega(\xi) \sigma_{ij} \varepsilon_{ij}^E =  \omega(\xi)\frac{1}{2}E(T)\varepsilon_{ij}^E\varepsilon_{ij}^E, \label{elastic free energy full}
\end{equation}
where $\omega(\xi)$ is the degradation function and $E(T)$ is the Young's modulus as a function of temperature. The degradation function $\omega$ describes how material strength changes with the damage phase field variable $\xi$ and is defined in Sec. \ref{phasefieldsection}.
 Finally, the total strain is given as a linear combination of the elastic strain $\varepsilon_{ij}^E$ and thermal expansion $\varepsilon_{ij}^T$; i.e.,
\begin{equation}
    \varepsilon_{ij} = \varepsilon^E_{ij} + \varepsilon^T_{ij},
\end{equation}
where the thermal strain $\varepsilon_{ij}^T$ of an isotropic material can be expressed in terms of the change in temperature $T$ and thermal expansion coefficient $\alpha$(T) as
\begin{equation}
    \varepsilon_{ij}^T = \alpha(T) \Delta T\delta_{ij},
\end{equation}
where $\delta_{ij}$ is the Kronecker delta.

\subsection{Phase Field} \label{phasefieldsection}
The damage phase field variable, $\xi$, is an auxiliary, non-conserved, scalar variable that regularizes the range between undamaged and damaged states; i.e.,
 \begin{equation}
     \xi(x_i) = \begin{cases}
        0, & \text{undamaged material;}\\
        (0,1), & \text{partially damaged material;}\\
        1, & \text{complete material failure.}
     \end{cases}
 \end{equation}
 The degradation function, $\omega(\xi)$, is a function of the phase field variable used to provide a continuous description of the diffusive damage zone. It is defined as 
\begin{equation}
   \omega(\xi) = (1-\xi)^2(1-\beta)+\beta, \label{degradation_full}
\end{equation}
with $\beta \ll 1$ enforcing numerical stability and has been introduced in several formulations \cite{Kuhn2015, Borden2012, Bourdin2008}.
The function satisfies the conditions
\begin{equation*}
    \omega(0) = 1,\qquad \omega(1)=0,
\end{equation*}
indicating that a non-damaged region ($\xi=0$) has the full material strength and a fully damaged region ($\xi=1$) can no longer store any elastic strain.

The evolution of $\xi$ is modeled using the Allen-Cahn phase field equation, which follows the variational principle of energy. First proposed by John W. Cahn and Sam Allen \cite{Allen1972}, the Allen-Cahn equation is defined as
\begin{equation}
    \label{Allen-Cahn}
    \frac{\partial \xi}{\partial t} = -L_{\xi} \left(\frac{\partial f_{loc}(\xi, \varepsilon^E_{ij})}{\partial \xi} - (\kappa \xi_{,i})_{,i}\right),
\end{equation}
where $L_{\xi}$ is the associated mobility constant, $f_{loc}$ is the local free energy, and $\kappa$ is the coefficient of the surface energy contribution from $\xi$.  The first component of the right-hand-side drives the system to a lower-energy state, while the second controls the sharpness of the transition between $\xi=0$ and $\xi=1$.

The local free energy is defined as a linear combination between the contribution from elasticity and the volumetric contribution from the damage phase field variable; i.e.,
\begin{equation*}
    f_{loc} = f_{elas} + f_{frac},
\end{equation*}
where $f_{elas}$ is the elastic free energy and $f_{frac}$ is the free energy from the phase field variable, $\xi$. 
Physically, once the material fractures, it can withstand compressive stress but not tensile stress. Therefore, it is assumed that the damage only affects the tensile strain and stress. Through strain spectral decomposition, the strain energy $\psi_{elas}$ is broken into tensile and compressive components $\psi^+$ and $\psi^-$; i.e.,
\begin{align*}
    \psi_{elas} &= \psi^+ + \psi^-.
\end{align*}
Since the crack does not self-heal, crack irreversibility is a necessary condition. To achieve this, a history variable $\mathcal{H}$ is introduced to record the maximum tensile strain energy throughout the time domain:
\begin{equation}
    \mathcal{H} = \max_t{(\psi^+, f_{barrier})},
\end{equation}
where
\begin{equation}
    f_{barrier} = -\frac{f_{frac}}{\omega(\xi)}\Big|_{\xi=0}.
\end{equation}
The free energy lower bound $f_{barrier}$ exists to ensure the tensile strain energy never becomes negative. Then the elastic free energy becomes 
\begin{equation*}
    f_{elas} = \omega(\xi)\mathcal{H}+\psi_{elas}^{\_}
\end{equation*}
Modeling the volumetric fracture energy term as a quadratic expression is a well-documented approach \cite{Miehe2010, Miehe2015}. The corresponding fracture free energy terms are then
\begin{equation}
    f_{frac} = \frac{g_c\xi^2}{2l_0}, \qquad \kappa = \frac{g_cl_0}{2}. \label{quad_frac_energy}
\end{equation}
where $g_c$ is the critical energy release rate defined by Griffith's theory of fracture \cite{Lajtai1971} and $l_0$ is the regularization crack length scale. The fracture energy formulation is presented explicitly in Sec. \ref{crack length scale}.

\subsection{Heat Conduction}
The heat conduction module is applied to account for the effects of temperature on material properties and on thermally induced expansion. It is also assumed that there is no heat generation within the material. {\color{black} There exist three modes of heat transfer: conduction, convection, and radiation. Convection is not possible within solid materials, but convection can be prescribe as thermal boundary conditions as well as radiation. Radiative heat transfer especially prominent at high temperatures, and it can be calculated by Stefan-Boltzmann law, $\Dot{Q}=\sigma_sA_s(T_{\infty}^4-T^4)$. In the case of an isotherm environment ($T=T_{\infty}$), no radiation heat transfer is needed. }The thermal conduction coefficient is coupled with the damage phase field variable, $\xi$, based on the assumption that there will be no conductive heat transfer once the material becomes totally damaged:
\begin{equation}
    \frac{\partial T}{\partial t} = \omega(\xi)\alpha(T) T_{,ii}. \label{heat conduction eq}
\end{equation}

\subsection{Coupling}
The coupling between the three modules is illustrated in Fig. 2.
The elasticity and phase field modules are coupled by letting the elastic free energy be a function of the damage phase field variable; i.e., $f_{elas} = f_{elas}(\xi)$.
The assumption that there is no heat transfer through totally damaged material allows the coupling between heat conduction and the damage phase field. With material properties as a function of temperature, the elastic strain energy term, after strain spectral decomposition, becomes
\begin{equation}
    \psi^\pm(T) =\frac{1}{2} \lambda_0(T)\langle\textstyle\sum\varepsilon_i\rangle_\pm^2\\ + \mu_0(T)(\textstyle\sum\langle\varepsilon_i\rangle_\pm)^2,
\end{equation}
where $\varepsilon_i$ are eigenstrains, $\lambda_0$ and $\mu_0$ are Lam\'e parameters given by
\begin{equation}
    \lambda_0(T) = \frac{\nu(T)E(T)}{(1+\nu(T))(1-2\nu(T))}, \qquad \mu_0(T) = \frac{E(T)}{2(1+\nu(T))},
\end{equation}
and the operator $\langle*\rangle_\pm = \frac{1}{2}(* \pm |*|)$.
This provides the complete links among the three modules.
The model is considered weakly two-way coupled, because temperature gradient is assumed to be negligible as heat conduction speed is much slower compared to the propagation speed of cracks.

\subsection{Crack Length Scale}\label{crack length scale}
The regularization crack length scale is vital in describing the free energy term contributed by the damage phase field variable as mentioned in Eq.~\ref{quad_frac_energy}. To avoid complex crack surface tracking, a diffusive crack modeling approach is adopted. Typical diffusive crack approximation takes on the form of a standard exponential function as shown in Fig. 3.
To determine an appropriate crack length scale the exponential function used to approximate the geometry of the diffusive crack is
\begin{equation}
    \xi_0 = e^{-|x|/l_0}. \label{xiexp}
\end{equation}
The crack length scale, $l_0$, determines the width of the diffusive zone of cracks. Physically, the smaller the crack length scale, the more accurate the results. However, numerical simulations have shown that too small of a crack length scale can result in a non-physical overshoot in material strength \cite{Zhang2017}. Therefore, the choice of $l_0$ significantly impacts the accuracy of the fracture model. Equation~\ref{xiexp} smears the sharp crack interface into a diffusive zone over the length of $l_0$ in both directions from the axial origin. The crack length scale is material dependent, as different materials have different fracture behaviors. A greater $l_0$ implies a larger diffusive zone. Cracks of brittle materials are usually smaller than those of ductile materials. The crack length scale significantly affects material strength in the simulation environment.

Based on the form of the crack surface geometry modeling, the exponential equation shown in Eq.~\ref{xiexp} is a solution to the ordinary differential equation
\begin{equation}
    -\xi''+\frac{1}{l_0^2}\xi = 0, \label{ode_gov}
\end{equation}
with boundary conditions
\begin{equation}
    \xi(0) = 1, \qquad \xi(\pm\infty) = 0.
\end{equation}
Integrating this ODE the regularized functional $A_d$ is defined as the weak form of Eq. \ref{ode_gov}:
 \begin{equation}
  A_d(\xi) \equiv \int_{-\infty}^\infty \left[\frac{\xi^2}{2l_0} + \frac{l_0|\nabla\xi|^2 }{2}\right] dV
 \end{equation}
 This aligns with a widely adopted approach which assumes that the volumetric free energy due to damage is a quadratic expression~\cite{Miehe2010, Molnar2020}.
 Based on the Griffith's theory of fracture, the energy required to create a new crack is equivalent to the amount of strain energy released. The critical energy release rate $g_c$ is defined to normalize such energy over the newly created surface area. Multiplying the functional with $g_c$ results in
\begin{equation}
    \Gamma = g_cA_d = \int_{-\infty}^\infty \left[\frac{g_c\xi^2}{2l_0} + \frac{g_cl_0|\nabla\xi|^2 }{2}\right] dV\label{ffrac},
\end{equation}
where $\Gamma$ represents both the volumetric and surface energy contributions of damaged materials, matching Eqs. \ref{Allen-Cahn}, \ref{quad_frac_energy}.

\subsubsection{Analytical Solution}
Any deformation tensor can be decomposed into symmetric and anti-symmetric parts, which are equivalent to tensile/compressive and shear contributions. Therefore, analytical solutions to these problems can be derived separately. These analytical solutions provide insight into how to determine the crack length scale used in a specific problem. Many have attempted to derive and propose the analytical solution to the crack length scale~\cite{Kuhn2015, Miehe2015, Zhang2017, Tanne2018}. The solution for two cases, simple tension and simple shear, are summarized here.

\subsubsection{Simple Tension} \label{simpletension}
Consider a homogeneous 1D weightless bar whose domain is $\Omega$, subjected to deformation and a phase damage parameter (cf. Fig. 4):
    \begin{equation}
        \sigma_{xx,x} = 0 \text{ in $\Omega$}, \label{gov_eqn1}
    \end{equation}
    \begin{equation}
        (1-\xi)\psi_{elas} - \frac{g_c}{2l_0}\xi + 2g_cl_0\xi_{,ii} = 0 \text{ in $\Omega$}, \label{gov_eqn2}
    \end{equation}
\noindent with boundary conditions
\begin{equation*}
    u_x(x=0) = 0, \qquad u_x(x=L) = \bar{u},
\end{equation*}
\begin{equation*}
    \frac{d \xi}{d x}|_{x=0} = 0, \qquad \frac{d \xi}{dx}|_{x=L} = 0.
\end{equation*}
Given that only tensile strain is present, the uncoupled elastic strain energy, $\psi_{elas}$, becomes
\begin{equation}
    \psi_{elas} = \frac{1}{2} E \varepsilon^2. \label{uncoupled elastic energy}
\end{equation}
For simplicity, the degradation function shown in Eq. \ref{degradation_full} can be reduced by eliminating the small constant $\beta$, i.e.,
\begin{equation}
    \omega(\xi) = (1-\xi)^2, \label{deg func simp}
\end{equation}
as $\beta$ does not carry any significance to the analytical derivations.
Multiplying Eq. \ref{deg func simp} by Eq. \ref{uncoupled elastic energy}, the same form of elastic free energy as in Eq. \ref{elastic free energy full} is recovered.
The strain of the bar is then
\begin{equation}
     \varepsilon = \frac{\bar{u}}{L}. \label{barstrain}
\end{equation}
Substituting Eq. \ref{barstrain} into Eq. \ref{uncoupled elastic energy} the elastic strain energy becomes
\begin{equation}
     \psi_{elas} = \frac{1}{2} E \varepsilon^2 = \frac{1}{2}\frac{E \bar{u}^2}{L^2},
\end{equation}
while the Cauchy stress becomes
\begin{equation}
    \sigma_{xx} = \omega(\xi)E \varepsilon = (1-\xi)^2\frac{E \bar{u}}{L}.  \label{tensile stress}
\end{equation}
The damage parameter can be obtained through Eq. \ref{gov_eqn2} along with the boundary conditions, resulting in
\begin{equation}
    \xi =  \frac{2(\frac{1}{2}E)l_0\Bar{u}^2}{2(\frac{1}{2}E)l_0\Bar{u}^2 + g_c L^2}. \label{tensile length scale}
\end{equation}
Now substituting Eq. \ref{tensile length scale} into Eq. \ref{tensile stress}, the stress is obtained as a function of $\Bar{u}$ only:
\begin{equation}
    \sigma_{xx}(\Bar{u}) = \frac{g_c^2 L^3 E \Bar{u}}{(2El_0\Bar{u}^2 + g_c L^2)^2}. \label{sigma_xx}
\end{equation}
This reveals the critical stress $\sigma_c$ by letting \(\frac{\partial\sigma_{xx}}{\partial \Bar{u}} = 0\).
The corresponding displacement $\Bar{u}^*$ is then
\begin{equation}
    \Bar{u}^* =\sqrt{\frac{g_c L^2}{3El_0}}. \label{u bar tensile}
\end{equation}
Substituting Eq. \ref{u bar tensile} into Eq. \ref{sigma_xx}, critical stress becomes
\begin{equation}
    \sigma_c = \frac{9}{16}\sqrt{\frac{E g_c}{3 l_0}}; \label{sigma c tensile}
\end{equation}
Rearranging yields expressions for the regularized crack length and the critical energy release rate:
    \begin{equation}
        l_0 = \frac{27}{256}\frac{g_c E}{\sigma_c^2} = \frac{27}{256} \frac{K_{Ic}^2}{\sigma_c^2}, \label{l_0}
    \end{equation}
    \begin{equation}
        g_c = \frac{K_{Ic}^2}{E} \label{tensile gc},
    \end{equation}
where $K_{IC}$ is the critical stress intensity factor under mode I fracture. Experimental data for \textit{E}, $\sigma_c$, and $K_{IC}$ are well-documented for many materials, from which $l_0$ and $g_c$ are straightforwardly derived through Eq. \ref{l_0} and Eq. \ref{tensile gc} and entered into the fracture free energy term shown in Eq. \ref{quad_frac_energy}.

\subsubsection{Simple Shear}
Besides simple tension, shear fracture constitutes another important part of fracture phenomena.
Consider a square subjected to a horizontal displacement, $\bar{u}$, at the top surface. The domain of interest is $\Omega$. The dimension of the square is set to be L$\times$L (cf. Fig. 5). The strain energy due to this shear deformation can be represented as
\begin{equation}
    \psi_{elas} = \sigma_{ij}\varepsilon^E_{ij} = \frac{1}{2}\mu\varepsilon_{ij}^E\varepsilon_{ij}^E.
\end{equation}
Let the ratio $\frac{\bar{u}}{L}$ be $\phi$; then, the elastic strain energy can be expressed as $\frac{1}{2} \mu \phi^2$. Since isotropy is assumed for the material, the shear direction is irrelevant to damage.
As before the governing equations involve deformation and a phase damage parameter:
    \begin{equation}
      \sigma_{ij,j} = 0 \text{ in $\Omega$}, \label{goveq3}
    \end{equation}
    \begin{equation}
        (1-\xi)\psi_{elas} - \frac{g_c}{2l_0}\xi + 2g_cl_0\xi_{,ii} = 0 \text{ in $\Omega$}, \label{goveq4}
    \end{equation}
with boundary conditions
\begin{equation*}
    u_x(x=0) = 0, \qquad u_x(x=L) = \bar{u},
\end{equation*}
\begin{equation*}
    \frac{d \xi}{d x}|_{x=0} = 0, \qquad \frac{d \xi}{dx}|_{x=L} = 0.
\end{equation*}
To solve Eq. \ref{goveq4}, a linear crack is assumed such that $\xi_{,ii}=0$, yielding
\begin{equation}
    (1-\xi)\psi_{elas} - \frac{g_c}{2l_0}\xi = 0,
\end{equation}
\begin{equation}
    \xi = \frac{l_0\mu\kappa^2}{l_0\mu\kappa^2  + g_c} = \frac{2l_0(\frac{1}{2}\mu)\Bar{u}^2}{2l_0(\frac{1}{2}\mu)\Bar{u}^2  + g_c L^2} . \label{xi simple shear}
\end{equation}
The Cauchy stress is then
\begin{equation}
     \sigma_{xy} = \omega(\xi)G \varepsilon_{xy}= (1-\xi)^2\frac{G \bar{u}}{L}. \label{stress shear}
\end{equation}
Eq. \ref{stress shear} is similar in form to that of Eq. \ref{tensile stress}. Substituting Eq. \ref{xi simple shear} into Eq. \ref{stress shear}, the stress is obtained as a function of $\Bar{u}$ only:
\begin{equation}
    \sigma_{xy}(\bar{u}) = \frac{g_c^2 L^3 (\frac{1}{2}\mu) \Bar{u}}{(2(\frac{1}{2}\mu)l_0\Bar{u}^2 + g_c L^2)^2} \label{shear stress bar}
\end{equation}
This allows us to find the critical stress by setting $\frac{\partial\sigma_{12}}{\partial \Bar{u}}=0$. The corresponding displacement $\bar{u}^*$ is
\begin{equation}
    \Bar{u}^* =\sqrt{\frac{g_c L^2}{3(\frac{1}{2}\mu)l_0}}. \label{bar star shear}
\end{equation}
Substituting Eq. \ref{bar star shear} into Eq. \ref{shear stress bar}, critical shear stress is then
\begin{equation}
    \sigma_{c,shear} = \frac{9}{16}\sqrt{\frac{(\frac{1}{2}\mu) g_c}{3 l_0}},
\end{equation}
yielding the properties of interest:
    \begin{equation}
        l_0 = \frac{27}{256}\frac{g_c (\mu)}{\sigma_{c,shear}^2} = \frac{27}{256} \frac{K_{IIc}^2}{\sigma_{c,shear}^2},
    \end{equation}
    \begin{equation}
        g_c = \frac{K_{IIc}^2}{\mu} \label{shear gc},
    \end{equation}
where $K_{IIC}$ is the fracture toughness under mode II, which is fracture under pure shear. Similar to the results shown for simple tension case, if given \textit{$\mu$}, $\sigma_c$, and $K_{IIc}$, the values for $l_0$ and $g_c$ are derived and entered into the fracture free energy term shown in Eq. \ref{quad_frac_energy}.
\subsubsection{Remark}
The analytical crack length scales are of very similar forms in both simple shear and simple tension. However, the fracture toughness data under pure shear are not as readily available as those in pure tension. It may be assumed that the critical energy release rate is identical for both mode I and mode II fractures. This allows for the estimation of $K_{II_C}$ by setting Eq. \ref{tensile gc} equal to Eq. \ref{shear gc}.

\section{Validation} \label{validation section}
The coupled framework is implemented in MOOSE, an open source finite element framework developed by Idaho National Lab \cite{Permann2020}.
The validation case for the proposed framework is flexural strength of $\alpha$-SiC. The temperature dependent material properties of $\alpha$-SiC are obtained from Munro~\cite{Munro1997}, which are also shown in Table \ref{tab:my-table}, where $u_k$ is 12$\%$ for T$\leq$400$^\circ$C and 8$\%$ for T$>$400$^\circ$C, and $\rho_0=3.16\pm0.03\ \text{g/cm}^3$ when $T_0=0 ^\circ \text{C}$ The mode I fracture toughness data are obtained from the straight notch bending test results reported in Ghosh et al~\cite{Ghosh1989}. As ceramics are generally too brittle to perform conventional tensile testing, bending tests are used to obtain a quantitative understanding of the critical strength properties. Flexural strength, by definition, is the maximum stress that occurs at the bottom surface of a specimen during bending tests. The bottom surface is where cracks first form due to tensile stress in these bending tests.  Since the tensile strength of brittle ceramics, like $\alpha$-SiC, are much less than the compressive strength of $\alpha$-SiC, the bottom surface will reach fracture faster than the top surface.


\subsection{Four Point Bending Test}
The numerical experiment is done through four point bending. The geometry is scaled down from the standard specified by ASTM C1161-18 \cite{C282013}. The mesh size is constrained in order to capture the crack geometry defined by the crack length scale. The general guideline is $h/l_0>1/10$, i.e. it requires at least 10 elements to capture the crack geometry.

Munro has collected four sets of bending tests on $\alpha$-SiC with various configurations \cite{Munro1997}. The general interpolation function of the flexural strength of $\alpha$-SiC is given as
\begin{equation}
    \text{FS (MPa)} = 359 + \frac{87.6}{1+208600 e^{-0.012T}} \pm 15\%
\end{equation}
{\color{black}Numerical simulations are conducted at 9 different temperatures, ranging from 20 to 1400 $^\circ$C in ambient air which is consistent with the experimental setup \cite{Munro1997}. The specimens are assumed to be held at the corresponding temperatures.} As the material properties used in the present model are measured in experiments, the uncertainty in the experimental properties will affect the simulation results. To explore the influence of this uncertainty the two extremes of the Young's modulus and Poisson's ratio, as shown in Table \ref{tab:my-table}, are used to create four sets of inputs. These inputs were then sent to the model to provide an estimate of the uncertainty of the output. The results are shown in Fig. 6. It shows that the region marked by light purple are still within the 15\% uncertainty bandwidth measured from experiments. {\color{black} There appears to be an increase and then decrease in predicted flexural strength between 400-1200$^\circ$C, which is similar to that of the experimentally measured critical energy release rate shown in Fig. 8. It should be noted that these experimentally obtained critical energy release rates are used in the phase field model for damage. However, uncertainties in other material properties that goes into the simulations also contribute to such changes. For example, changes in flexural strength over this temperature range are comparable to the changes by varying Young's modulus and Poisson's ratio.}

From the simulation results, one can see that the the flexural strengths are well aligned with the expected experimental data outside the transition region, 800$^\circ$C to 1200$^\circ$C . This behavior of increasing flexural strength as temperature increases in the transition region is because oxidization on the surface of the specimen results in short-term crack healing. This chemical phenomena, however, is not incorporated in the current framework. Once the temperature passes the transition region{\color{black}, an oxidization layer is formed on the surface of the material.} When the newly formed oxidation layer is taken into account as the bulk property {\color{black} 
 such as Young's modulus and Poisson's ratio that are also measured at elevated temperatures}, the flexural strength is once again well aligned with the expected curve.

\subsection{Fracture Toughness}
In linear elastic fracture mechanics (LEFM), one of the classic examples is an infinite plate under uniform stress. Fracture Toughness, $K$, is defined as the stress intensity factor when the material's crack starts to propagate.
Sneddon's solution provides analytically-derived displacements and stress responses for any point in the domain, for an infinite plate under uniform stress \cite{Sneddon1969, Chen2015}. If one applies the Sneddon's solution of displacement responses, shown in Eqs.~\ref{snedon_ux} and~\ref{snedon_uy} as the boundary condition to a finite domain with a centered crack, it essentially recreates the same problem as an infinite plate. Fig. 7 shows the problem setup of Sneddon's solution.
    \begin{equation}
        \begin{split}
           4\mu u_x &= \frac{K_1}{\sqrt{arr_2}}\left[ (\zeta - 1)r r_2\cos{\frac{\theta + \theta_2}{2}} - 2r_1^2\sin{\theta_1}\sin{\left( \theta_1 - \frac{\theta + \theta_2}{2}\right)}\right]\\
            &+\frac{K_2}{\sqrt{arr_2}}\left[ (\zeta + 1)r r_2\sin{\frac{\theta + \theta_2}{2}} + 2r_1^2\sin{\theta_1}\cos{\left( \theta_1 - \frac{\theta + \theta_2}{2}\right)}\right] \\
            &- 0.5r_1\cos{\theta_1}(\sigma-\chi)(\zeta + 1)
        \end{split}{} \label{snedon_ux}
    \end{equation}{}
    \begin{equation}
        \begin{split}
           4\mu u_y &= \frac{K_1}{\sqrt{arr_2}}\left[ (\zeta + 1)r r_2\sin{\frac{\theta + \theta_2}{2}} - 2r_1^2\sin{\theta_1}\cos{\left( \theta_1 - \frac{\theta + \theta_2}{2}\right)}\right]\\
            &+\frac{K_2}{\sqrt{arr_2}}\left[ (1 - \zeta)r r_2\cos{\frac{\theta + \theta_2}{2}} - 2r_1^2\sin{\theta_1}\sin{\left( \theta_1 - \frac{\theta + \theta_2}{2}\right)}\right] \\
            &- 0.5r_1\sin{\theta_1}(\sigma-\chi)(\zeta - 3)
        \end{split}{}\label{snedon_uy}
    \end{equation}{}
    where
    \begin{equation}
        K_1 = \sigma\sqrt{ a} \qquad K_2 = \tau \sqrt{ a}
    \end{equation}
    \begin{equation}
        \zeta = \begin{cases}
            3-4\nu & \qquad \text{for plane strain,}\\
            \frac{3-\nu}{1+\nu} & \qquad \text{for plane stress,}
        \end{cases}
    \end{equation}
and a is the half crack length. By specifying the values of the far-field stresses, $\sigma,\ \tau,\ \chi$, one can make the problem to mode I, II or mixed mode.

\subsubsection{Mode I}
Mode I fracture refers to the case when the material is under simple tension. The simulations utilize a square domain and Sneddon's displacement response under far field tensile stress, $\sigma$, as boundary conditions. The mesh size is uniform 120$\times$120 cells with dimensions of 0.3$\times$0.3 mm. The half center crack length, $a$, is 25 $\mu$m.
Fig. 8 shows that the general trend of the fracture toughness obtained from simulations are well aligned with the experimental data, with the the results from our simulations being on the lower end of the confidence interval. It is worth noting that the experimental data {\color{black} of mode I fracture toughness }are obtained from {\color{black}single edge notched bending (SENB)} specimens with straight notches and that tests with different {\color{black} size and shapes of the }notches are known to lead to different results. The numerical results, although at most 10.9\% lower than the experimental data, are still within the acceptable range.

\subsubsection{Mode II}
Mode II fracture refers to the case when the material is under simple shear. The simulations utilize a square domain and Sneddon's displacement response under far field shear stress, $\tau$, as boundary conditions. It can be seen that the model II fracture toughness decreases slightly as temperature increases, which also shows that fracture toughness for mode II is not very sensitive to temperature. This is consistent with literature which shows that the fracture toughness of $\alpha$-SiC under mode I fracture is temperature independent, while data for mode II fracture toughness of $\alpha$-SiC is relatively independent \cite{Ghosh1989}. {\color{black}There are not sufficient data for mode II fracture toughness of $\alpha$-SiC over the proposed wide range of temperatures from 20 to 1400 $^\circ$C. Since the predicted mode I fracture toughness data have been validated, one can infer that the predicted mode II fracture toughness data that follows the same set of definitions are reliable.}

\subsection{Modified G-Criterion}
In general, materials fracture under both shear and tensile stress.
Experiments have suggested the following modified G-criterion \cite{Sun2012}:
\begin{equation}
    \left(\frac{K_I}{K_{I_C}}\right)^2 + \left(\frac{K_{II}}{K_{{II}_C}}\right)^2 = 1
\end{equation}
where
\begin{equation}
    K_{IC} = \sigma_{C}\sqrt{\pi a} \qquad  K_{IIC} = \tau_{C}\sqrt{\pi a}
\end{equation}
and $\sigma_{C},\ \tau_{C}$ are the corresponding critical far field stresses when a crack first appears.
A total of eight temperatures ranging from 0$^{\circ}$C to 1400$^{\circ}$C are considered, with numerical results for three of these temperatures provided in Table \ref{critical stress intensity}.
In this table case 1 is mode I, case 2 is mode II, and cases 3-5 are mixed modes.
Fig. 10 shows that the fracture toughness under mixed mode fracture is generally on the lower side, consistent with the findings for Mode I fracture toughness. As the simulation results of mode II fracture toughness are more consistent that those of mode I fracture toughness, the profile of modified G-criterion is mainly affected by mode I fracture fracture toughness. This provides a more realistic guideline for engineering designs. Finally, Fig. 11 shows that the proposed framework is capable of modeling the fracture phenomena under various stress modes and temperatures. Similar to Fig. 9, the mixed mode fractures also show relative temperature independence.

\section{Scalability} \label{scallability section}

Work in MOOSE is undergirded by a library of numerical solvers offered by PETSc, a widely used parallel high performance computing package \cite{PETSc2022-1, PETSc2022-2, PETSc2022-3}. As outlined by Chang et al \cite{Chang2018}, the following scalability tests are motivated by answering the question of which iterative methods offered by PETSc are best for {\color{black}solving} the present set of partial differential equations across various domain sizes, degrees of freedom, and computer architectures.

\subsection{Solver Choice}
Information on scalability of a solver-preconditioner combination, along with a procured set of associated parameters, is only relevant after it is clear that such a solver converges to a sufficiently accurate result. Therefore, the prerequisite criteria to scalability testing are stability, i.e. convergence, {\color{black}and} reliability, i.e. accuracy with respect to a direct solve. The mesh used to test these is a coarse two-dimensional square with a free surface established at the middle of the left edge as a demonstration of mode I fracture. The length of each side of a cell relative to the physical domain size is $\Delta x = \Delta y = 10^{-2}$. The resolution of the mesh is low to make possible a direct calculation via the Newton method, the result with which error analysis can be performed in comparison to a particular iterative solver-preconditioner combination. Solver options considered are the Jacobian-free Newton-Krylov method (JFNK) and preconditioned JFNK (PJFNK); preconditioning options considered are LU decomposition (LU), the Additive Schwarz method (ASM), and an algebraic multigrid method (BoomerAMG). PETSc offers a suite of customizations associated with the BoomerAMG preconditioner, such as matrix entry truncation and the intensity of aggressive coarsening. Various combinations of these sub-parameters are also considered.
\subsubsection{Convergence}
The assembler output file generated by MOOSE returns in a message whether or not the residual has fallen below the preset nonlinear relative tolerance after a number of time steps less than what captures the onset of crack propagation. Such a time window is considered as opposed to that of crack propagation itself because the probability of numerical divergence following material failure is much higher and would not be indicative of solver reliability.
\subsubsection{Accuracy}
For a particular test, after {\color{black}testing} for convergence, the average $L_2$ norm of strain components $\varepsilon_{xx}, \ \varepsilon_{yy}$ in the mesh is subject to the Newton method among $n$ cells according to
\begin{equation}
    \text{error per cell = }\frac{\sum_{n} \sqrt{\delta \varepsilon_{xx}^2 + \delta \varepsilon_{yy}^2}}{n}. \label{l2}
\end{equation}
The average $L_2$ norm is used as a metric for computational accuracy of the iterative solver-preconditioner combination {\color{black}of} that test.
\subsubsection{Result}
Following the prescribed testing, a PJFNK-BoomerAMG variant with a low setting on aggressive coarsening emerged as the optimal combination for the current coupled scheme on the basis of cell accuracy, with a difference of $1.06\times10^{-3}$ per cell.  This difference in cell accuracy is comparable to the leading term of the finite difference approximation of the displacement derivative $\mathcal{O}(\Delta x^2) \sim 10^{-4} = \Delta x^2$.

\subsection{Scaling methods}
The PJFNK-BoomerAMG combination is chosen for scalability testing, a procedure divided into strong scaling, weak scaling, and static scaling.
\subsubsection{Strong scaling}  \label{strong}
Several simulations with different numbers of cores attributed to a fixed domain are measured on wall clock time in strong scaling. By finding the job time of $N$ cores, $t_N$, the parallel proportion of the input file $p \in [0,1]$ is recovered according to a rearranged Amdahl's Law,
\begin{equation}
    p = \frac{1-t_N/t_1}{1-1/N} = \frac{1-1/S(N)}{1-1/N}, \label{amdahl}
\end{equation}
where a higher $p$ indicates a more scalable algorithm and $S(N)=t_N/t_1$ indicates the speedup of using $N$ cores compared to the 1-core case. Fig. 12 is a visualization of the mesh used, with differences in color representing discrete sub-domains, and Fig. 13 maps algorithmic speedup as a function of the number of cores utilized. From this we obtain $p=0.862$, {\color{black}which represents the average of the data set}.



\subsubsection{Weak scaling}  \label{weak}
In weak scaling, wall clock time is measured again, but now the number of degrees of freedom per core is fixed while the size of the domain is increased. In test $i$, for an $n$-dimensional mesh with square faces ($n \geq 2$), with $C$ cells per interval, $P$ characteristic PDEs, and domain scale $D$, degrees of freedom is given by
\begin{equation*}
    \text{DoF}_i = P[CD_i^{1/n}+1]^n.
\end{equation*}
In the current scheme we take $P=5, n=2, C=200$, and $D = \{4^2, 8^2, 16^2\}$. Weak scaling is predicated on the relationship
\begin{equation}
    \text{DoF}_i/N_i = \text{constant};
\end{equation}
therefore,
\begin{equation}
    \text{DoF}_i/N_i = \text{DoF}_1/N_1 = \frac{5[200(16)^{1/2}+1]^2}{1} = 3,208,005
\end{equation}
provided the control $N_1 = 1$. Locating the appropriate $N$ for a chosen $D$ requires
\begin{equation*}
    N_i = \frac{\text{DoF}_i}{\text{DoF}_1/N_1} = \frac{P[CD_i^{1/n}+1]^n}{P[CD_1^{1/n}+1]^n/1} \approx \frac{P[CD_i^{1/n}]^n}{P[CD_1^{1/n}]^n} = \frac{D_i}{D_1},
\end{equation*}
giving simply
\begin{equation}
    N_i \approx \frac{D_{i}}{D_{1}}.
\end{equation}
Parallelizability $p$ is again recovered, now in weak scaling according to a rearranged Gustafson's Law,

\begin{equation}
    p = \frac{S_s(D)-1}{N(D)-1} \label{gustafson}.
\end{equation}

\noindent Fig. 14 is a visualization of the meshes used, and Fig. 15 maps scaled algorithmic speedup as a function of the number of cores utilized.

The procedure yields $p = 0.463$, meaning that the case is moderately parallelizable subject to an increased domain size, although less so relative to the strong scaling metric.

\subsubsection{Static scaling} \label{static}
Static scaling subsumes features of both strong and weak scaling, while also offering an optimal scaling region in units of degrees of freedom per core \cite{Chang2018}. In strong scaling, complex computer architectures managing a relatively small problem size experience latency rooted in communication inefficiency. This is shown in the right-most experimental data points on Fig. 13. Also, in weak scaling, given that the number of degrees of freedom exponentially increases between tests, failure of the experimental data points in Fig. 15 to maintain a linear shape can be attributed to memory overloading effects. Static scaling experimentation across tests only changes the number of cells per interval, keeping constant the domain size ($4^2$) and the number of CPUs (2 nodes $\times$ 32 cores per node $=$ 64 cores $=N$). DoF per job time is plotted against the same job time in Fig. 17 to reveal algorithmic speed, under the assumption that each degree of freedom carries equal magnitude. The PJFNK solver continually used in the present scalability testing is compared to a direct Newton method to demonstrate the utility of an iterative method for large problem sizes. Ten tests are repeated for each case to ensure procedural reproducibility. Fig. 16 illustrates the various mesh discretizations.



As shown by Fig. 17, the combined PJFNK-BoomerAMG method is more scalable than the Newton method by all definitions. {\color{black}Therefore, this solver is chosen for the numerical experiments shown in Sec. \ref{validation section}.} Whereas the optimal scaling region for the Newton method is narrow, that of PJFNK-BoomerAMG spans the rest of the domain after $C \geq 100$ which corresponds to $\text{DoF}/N \geq 12,563$. Above this lower bound, communication costs are modest. At the upper bound, memory overloading effects are theoretically demonstrated, but no such bound was located in the set of problem sizes investigated, indicating that even for $C=375$ cells-per-interval with a DoF per node of 176,015, which corresponds to a total number of DoF of 11,265,005, memory effects are negligible. An optimal scaling region of at least $12,563 \leq \text{DoF}/N \leq 176,015$ implies the capability of diverse problem {\color{black}sizes} to be solved with high parallel efficiency.

\section{Conclusion}
A framework that models ceramic in a wide range of temperatures that couples elasticity, fracture phase field, and heat transfer has been developed. Four-point bending tests showed agreement of the flexural strengths predicted by the model's confidence intervals. Mode I fracture tests of infinite plate showed that the fracture toughness of mode I were also well matched with the experiments. When combined with mode II fracture, the modified G-criterion provided a general guideline for application design under mixed mode fractures. The simulation results showed independence of temperatures for various modes of fractures across a wide temperature range. These validation tests proved the reliability for damage prediction, linking between performance and material properties as detailed in the three-link ICME model.
The uncertainty quantification presented in this study, while still preliminary, revealed that flexural strength is not qualitatively influenced by the uncertainty of the material properties. At temperature ranges from 800-1200$^\circ$C, the oxidization effect that causes short-term crack healing has not yet considered in this framework.
It is noteworthy that the fracture tests were based on an analytical infinite plate case. The geometric difference between the standard notch tests and theoretical infinite plate contribute much of the disparity. Scalability tests, following rigorous convergence and accuracy checks, were also conducted on the iterative solver-preconditioner combination functioning as the numerical solver. Results demonstrated that using the proper solver results is scalable, allowing for a wide range of problem sizes which can be solved. Future investigation will be focusing on adding the microstructure model, which will link manufacturing processes and material properties to the system performance, such that a complete ICME model for advanced ceramics can be used to facilitate a complete design to production cycle.

\section{Acknowledgement}
The authors wish to acknowledge their appreciation for their frequent constructive discussions with Dr. David Hicks, Dr. Amberlee Haselhuhn, and Bradley Friend from LIFT. Funding for this project was provided, in part, by LIFT, the Detroit-based national manufacturing innovation institute operated by ALMMII, the American Lightweight Materials Manufacturing Innovation Institute, a Michigan-based nonprofit, 501(c)3 as part of ONR Grant N00014-21-1-2660.

\bibliography{bib}
\bibliographystyle{vancouver}


\newpage
\section{Figure/table Caption List}
\setstretch{2}
\textbf{Figure 1}: The linear hierarchical structural relation of material science and engineering.\\
\textbf{Figure 2}: Schematic of elasticity, phase field, and heat conduction coupling.\\
\textbf{Figure 3}: Sharp and diffusive crack topologies. (a) Sharp (b) Diffusive.\\
\textbf{Figure 4}: Schematic of simple tension problem.\\
\textbf{Figure 5}: Schematic of simple shear problem.\\
\textbf{Figure 6}: Simulation and experimental flexural strength comparison of $\alpha$-SiC at various temperatures \cite{Munro1997}. Uncertainty bands are shown as well.\\ 
\textbf{Figure 7}: Geometry and loading conditions of infinite plate with center crack.\\
\textbf{Figure 8}: Mode I critical stress intensity factor at various temperatures are plotted along side with the experiment data.\\
\textbf{Figure 9}: Mode I and mode II stress intensity factors versus temperature.\\
\textbf{Figure 10}: Stress intensity factors of mixed mode fracture with the modified G-criterion model for (a) T = 20$^\circ$C; (b) T = 600$^\circ$C; (c) T = 1400$^\circ$C.\\
\textbf{Figure 11}: Isosurfaces for modified G-criterion at temperatures from 20$^\circ$C to 1400$^\circ$C.\\
\textbf{Figure 12}: 1200$^2$ element mesh used in strong scaling (300 cells per interval).\\
\textbf{Figure 13}: Speedup S = t1/tN versus number of CPUs N.\\
\textbf{Figure 14}: Mesh series used in weak scaling (200 cells per interval). (a): D = 4$^2$; (b): D = 8$^2$; (c): D = 16$^2$. \\
\textbf{Figure 15}: Scaled speedup versus number of CPUs.\\
\textbf{Figure 16}: Mesh series used in static scaling (variable cells per interval).\\
\textbf{Figure 17}: DoF/job time versus job time.\\
\textbf{Table 1}: Experimental Material Properties as Functions of Temperature.\\
\textbf{Table 2}: Fracture Toughness Under Mixed Modes at T = 20, 600, 1400 $^\circ$C.

\newpage
\section{List of Tables}
\begin{table}[!ht]
\centering
\caption{Experimental Material Properties as Functions of Temperature.}
\label{tab:my-table}
\resizebox{0.8\columnwidth}{!}{%
\begin{tabular}{|c|c|}
\hline
                          & Function                                                                             \\ \hline
$E (GPa)    $               & 415-0.023T$\pm\ 3\%$                                                                    \\ \hline
$\nu$                       & $0.160 - 2.62\cdot 10^{-6}T\pm 25\% $                                                     \\ \hline
$C_p$ $(Jkg^{-1}K^{-1})$  & $1110+0.15\cdot T-425e^{-0.003T} \pm 5\%$                                                 \\ \hline
$k (Wm^{-1}K^{-1})$       & $52000 e^{-1.24\cdot 10^{-5}}(T+219)^{-1}\pm u_k$                                         \\ \hline
$\rho (g/cm^3)$           & $\rho_0 [1+(4.22+8.33\cdot 10^{-4}T - 3.51e^{-0.00527T} \pm 10\%)(T-T_0)]^{-3}$               \\ \hline
\end{tabular}%
}
\end{table}\noindent

\begin{table}[!ht]
\centering
\caption{Fracture Toughness Under Mixed Modes at T = 20, 600, 1400 $^\circ$C.}
\label{critical stress intensity}
\resizebox{\columnwidth}{!}{%
\begin{tabular}{|c|c|c|c|c|c|c|}
\hline
Temperature ($^\circ$C) &
  \begin{tabular}[c]{@{}c@{}}Stress Intensity Factor\\ (MPa$\cdot \text{m}^{\text{1/2}}$)\end{tabular} &
  \begin{tabular}[c]{@{}c@{}}Case 1\\ ($\tau$ = 0)\end{tabular} &
  \begin{tabular}[c]{@{}c@{}}Case 2\\ ($\sigma$ = 0)\end{tabular} &
  \begin{tabular}[c]{@{}c@{}}Case 3\\ ($\sigma$ = 2$\tau$)\end{tabular} &
  \begin{tabular}[c]{@{}c@{}}Case 4\\ ($\sigma$ = $\tau$)\end{tabular} &
  \begin{tabular}[c]{@{}c@{}}Case 5\\ ($\tau$ = 2$\sigma$)\end{tabular} \\ \hline
\multirow{2}{*}{20}   & K$_{\text{IC}}$  & 3.4845 & 0      & 2.9583 & 2.2229 & 1.3820 \\ \cline{2-7}
                      & K$_{\text{IIC}}$ & 0      & 3.3912 & 1.4791 & 2.2229 & 2.7640 \\ \hline
\multirow{2}{*}{600}  & K$_{\text{IC}}$  & 3.6233 & 0      & 2.9102 & 2.1865 & 1.3593 \\ \cline{2-7}
                      & K$_{\text{IIC}}$ & 0      & 3.3317 & 1.4551 & 2.1865 & 2.7186 \\ \hline
\multirow{2}{*}{1400} & K$_{\text{IC}}$  & 4.2452 & 0      & 2.8432 & 2.1357 & 1.3272 \\ \cline{2-7}
                      & K$_{\text{IIC}}$ & 0      & 3.2528 & 1.4216 & 2.1357 & 2.6544 \\ \hline
\end{tabular}%
}
\end{table}

\newpage

\end{document}